\documentclass[sigconf,nonacm]{acmart}

\PassOptionsToPackage{table,xcdraw}{xcolor} 
\usepackage{xcolor} 

\usepackage{dirtytalk} 
\usepackage{wrapfig}
\usepackage{subcaption}
\usepackage[bottom]{footmisc} 
\usepackage{booktabs}
\usepackage{graphicx}
\usepackage{multirow}
\usepackage{pifont}
\usepackage{colortbl}

\newcommand{\sayit}[1]{{\textit{\say{#1}}}}

\usepackage{hyperref}
\usepackage{multirow}
\usepackage{colortbl}
\usepackage{booktabs}
\usepackage{graphicx}
\usepackage{soul}
\usepackage[table,xcdraw]{xcolor}
\makeatletter
\renewcommand{\footnoterule}{%
  \kern -3pt
  \hrule width 0.25\columnwidth
  \kern 2.6pt}

\long\def\@makefntext#1{%
  \parindent 0pt%
  \leftskip 0pt%
  \noindent
  $^{\@thefnmark}$~#1}
\makeatother

\newcommand{\richard}[1]{{\textcolor{black}{#1}}}
\newcommand{\xia}[1]{{\textcolor{black}{#1}}}

\copyrightyear{2025}
\acmYear{2025}
\setcopyright{cc}
\setcctype{by}
\acmConference[UIST '25]{The 38th Annual ACM Symposium on User Interface
Software and Technology}{September 28-October 1, 2025}{Busan, Republic of
Korea}
\acmBooktitle{The 38th Annual ACM Symposium on User Interface Software and
Technology (UIST '25), September 28-October 1, 2025, Busan, Republic of
Korea}
\acmDOI{}
\acmISBN{}

\makeatletter
\renewcommand{\paragraph}{%
  \@startsection{paragraph}{4}%
  {\z@}{0.60ex \@plus 1ex \@minus .15ex}{-1em}%
  {\normalfont\normalsize\bfseries}%
}
\makeatother

\begin{document}

\title{FlyMeThrough: Human-AI Collaborative 3D Indoor Mapping \\ with Commodity Drones}



\author{Xia Su}
\authornote{Both authors contributed equally to this research.}
\affiliation{%
  \institution{University of Washington}
  \city{Seattle}
  \state{Washington}
  \country{United States}
}
\email{xiasu@cs.washington.edu}

\author{Ruiqi Chen}
\authornotemark[1]
\affiliation{%
  \institution{University of Washington}
  \city{Seattle}
  \state{Washington}
  \country{United States}
}
\email{ruiqich@uw.edu}

\author{Jingwei Ma}
\affiliation{%
  \institution{University of Washington}
  \city{Seattle}
  \state{Washington}
  \country{United States}
}
\email{jingweim@cs.washington.edu}

\author{Chu Li}
\affiliation{%
  \institution{University of Washington}
  \city{Seattle}
  \state{Washington}
  \country{United States}
}
\email{chuchuli@cs.washington.edu}

\author{Jon E. Froehlich}
\orcid{0000-0001-8291-3353}
\affiliation{%
  \institution{University of Washington}
  \city{Seattle}
  \state{Washington}
  \country{United States}
}
\email{jonf@cs.washington.edu}
\renewcommand{\shorttitle}{FlyMeThrough}
\begin{abstract}
Indoor mapping data is crucial for routing, navigation, and building management, yet such data are widely lacking due to the manual labor and expense of data collection, especially for larger indoor spaces. Leveraging recent advancements in commodity drones and photogrammetry, we introduce \textit{FlyMeThrough}---a drone-based indoor scanning system that efficiently produces 3D reconstructions of indoor spaces with human-AI collaborative annotations for key indoor points-of-interest (POI) such as entrances, restrooms, stairs, and elevators. We evaluated FlyMeThrough in 12 indoor spaces with varying sizes and functionality. To investigate use cases and solicit feedback from target stakeholders, we also conducted a qualitative user study with five building managers and five occupants. Our findings indicate that \textit{FlyMeThrough} can efficiently and precisely create indoor 3D maps for strategic space planning, resource management, and navigation.

\end{abstract}

\ccsdesc[500]{Human-centered computing~Interactive systems and tools}
\ccsdesc[500]{Human-centered computing~Interaction design}
\ccsdesc[500]{Human-centered computing~Visualization systems and tools}

\keywords{Drone; 3D reconstruction; Indoor Mapping; Video Segmentation; Human-AI Collaboration}

\begin{teaserfigure}
  \includegraphics[width=\textwidth]{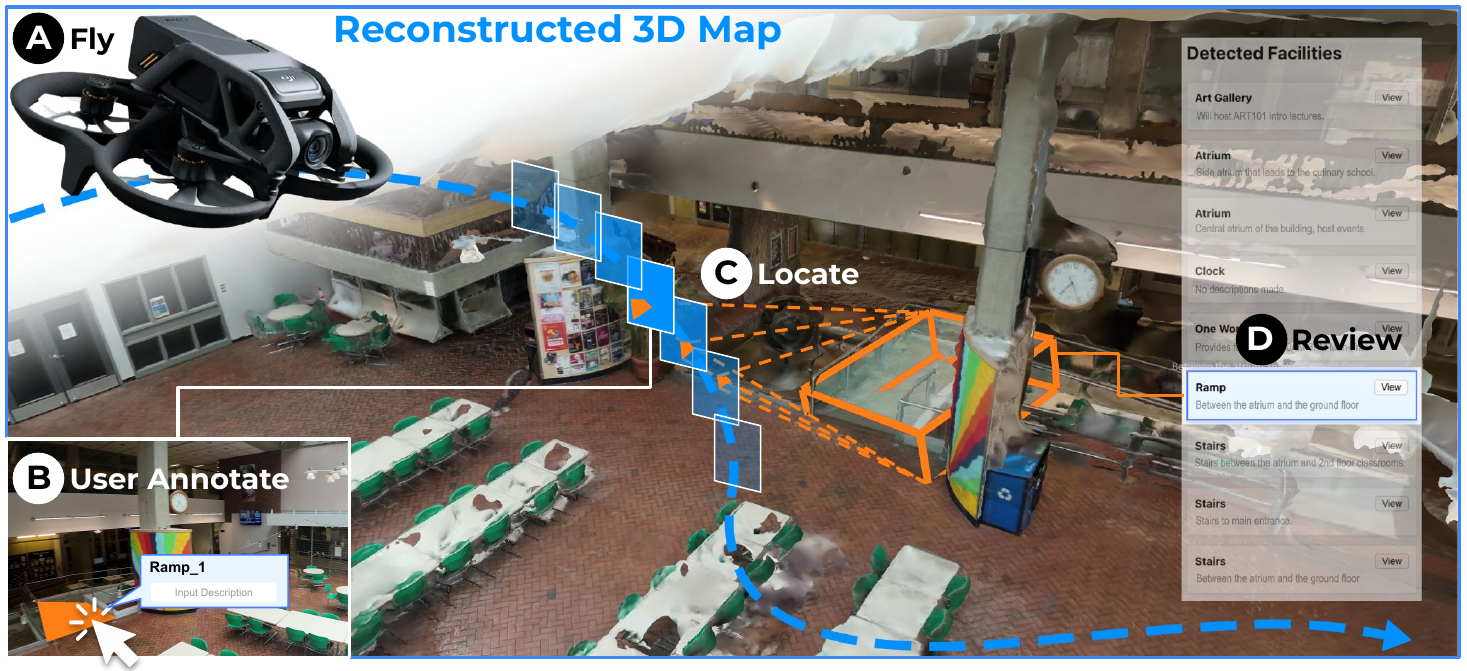}
  \caption{We introduce \textit{FlyMeThrough}, a drone-based indoor mapping system that semi-automatically maps indoor spaces and locates key facilities such as entrances, stairs, elevators, and doors. (A) We use a \textit{DJI Avata} drone to scan indoor spaces, and generate reconstructed 3D maps with structure-from-motion (SfM)~\cite{ozyecsil2017survey}; (B) We employ \textit{SAM2}~\cite{sam2_github} to enable intuitive and efficient user annotation of key indoor facilities. (C) FlyMeThrough auto-segments objects, generates annotations across subsequent frames, and localizes them in the 3D map, which (D) users can interactively review and update.}
  \Description{A drone flying and scanning an indoor space. The image also shows the indoor space, which is a 2-floor-high hall space, as a 3D reconstruction, since some missing geometry can be seen. The drone has a blue dotted flying trajectory, where several image frames are located on. One frame is expanded on the left bottom corner, showing a user annotation action that clicks the screen and gets a ramp highlighted with orange. Two more subsequent frames in the frame list also highlight this orange block. All these frames are linked to an orange bounding box with orange dotted lines, indicating that these masks on images cast to a 3D bounding box in the scene, which is linked to a "ramp" item in a list of detected facilities.}
  \label{fig:teaser}
\end{teaserfigure}

\maketitle

\section{Introduction}

Indoor mapping data for public indoor spaces (\textit{e.g.}, office buildings, train stations, airports, and stadiums, \textit{etc.}) is crucial for navigation, route planning, space evaluation, and tracking spatial changes over time. However, such data is often scarce and outdated \cite{pipelidis2019bootstrapping}. Mainstream indoor mapping procedures involve two methods: transforming CAD building plans into digital indoor mapping, or deploying specific scanning hardware and services. However, these methods require high data acquisition costs and long timelines \cite{yue2018update} and thus are hard to mass deploy and regularly maintain. 



In contrast, drone-based indoor mapping, where drones fly through \sloppy large indoor spaces to capture key spatial information, offers advantages in terms of cost and maintenance efforts. Demonstrated feasible through both academic research \cite{karam2022microdrone,zhou2014embedded} and commercial products like \textit{Skydio} \cite{skydio_3d_indoor_capture} and \textit{Elios} \cite{elios3_flyability}, drone-based mapping can efficiently cover large and complex indoor spaces and has the potential for higher levels of automation. However, existing solutions often rely on LiDAR-equipped drones, which are prohibitively expensive for large-scale deployment (\textit{e.g.}, an \textit{Elios 3} LiDAR drone costs \$50,000). Additionally, the generated indoor maps lack points of interest (POI) that building management and visitors care about. To address these challenges, we aim to develop a more affordable and scalable drone-based indoor mapping pipeline that only uses RGB video data, thus can potentially be captured by any commodity drone. In addition, we introduce a human-AI collaborative annotation and location process that allows building management staff to efficiently identify and mark key indoor POIs (\textit{e.g.} entrances, stairs, elevators), making the output 3D maps more useful for navigation, routing, and space evaluation.

We present \textit{FlyMeThrough}, a human-AI collaborative indoor mapping system that provides an end-to-end pipeline for transforming RGB footage of indoor drone flights to POI-infused 3D reconstructed maps of the scanned indoor spaces. FlyMeThrough is comprised of three major technical components: first, a SfM (Structure from Motion)-based 3D reconstruction \cite{ozyecsil2017survey} that transforms input RGB videos to estimated camera positions of video frames as well as 3D mesh models of the indoor spaces. Second, a human-AI collaborative annotation pipeline that enables users to efficiently annotate and locate key indoor POIs. Third, a web interface that reviews the final results, which are 3D models with bounding boxes indicating the locations and dimensions of key indoor POIs. All components of the system are released and open-sourced.\footnote{\url{https://github.com/makeabilitylab/FlyMeThrough}}


To evaluate FlyMeThrough, we collected drone footage from 12 indoor spaces of varying sizes and functionalities under the approval and guidance of building managers, who are responsible for maintaining these scanned indoor spaces. We also conducted a user study with these five building managers along with five active building occupants of these scanned spaces, to assess the performance and usefulness of the annotation interface. We presented our drone-based indoor mapping process, invited participants to annotate the collected data for key indoor POIs, then interviewed them to assess the usability and performance of our custom pipeline. The study results show high technical performance and usability of our mapping system, while revealing application scenarios and aspects of future improvements.


\richard{In sum, our contributions are threefold: first, we designed and implemented the first drone-based indoor mapping system that leverages only RGB data---enabling the use of affordable, off-the-shelf consumer drones for large-scale indoor reconstruction. Second, we introduce a custom human-AI collaborative annotation pipeline that allows users to flexibly define and create POIs in reconstructed indoor 3D maps. Finally, our user study with building managers and occupants contributes key insights into practical needs, usability, and potential applications of drone-based indoor 3D mapping systems for future work. }

\section{Related Work}
We situate our work in research on large-scale indoor mapping, drone-based 3D reconstructions, and techniques to automatically detect and localize objects in scenes.

\subsection{Large Space Indoor Mapping}
Large indoor spaces, such as train stations, airports, malls, and office buildings, require high-quality spatial maps to support applications like navigation, space management, and digital twin construction. Mainstream commercial indoor mapping services like \textit{Pointr} \cite{pointr}, \textit{ESRI} \cite{arcgis_indoors}, \textit{Mappedin} \cite{mappedin}, and \textit{Mapsted} \cite{mapsted} typically rely on existing or user-drawn architectural data as input to generate interactive indoor navigation and management systems. Meanwhile, 3D indoor scanning services like \textit{Matterport} \cite{noauthor_capture_nodate}, \textit{Cupix} \cite{cupix}, and \textit{NavVis} \cite{navvis} utilize LiDAR-equipped professional devices to provide high-precision 3D reconstruction and virtual visualization. In recent years, applications like \textit{PolyCam} \cite{polycam} and \textit{LumaAI} \cite{lumaai} have explored the creation of digital twins for small to medium scale indoor spaces using mobile LiDAR or RGB videos, while APIs such as Apple's \textit{RoomPlan} \cite{apple_roomplan_2022} further lower the threshold for small-scale 3D scanning. Although these commercial solutions are technologically mature, they still exhibit significant limitations. On the one hand, their reliance on professional hardware or manual data collection leads to high costs for data acquisition and subsequent updates. On the other hand, mobile-based solutions are typically designed for small indoor spaces and cannot efficiently scale to large environments spanning thousands of square meters.

In contrast to commercial approaches, recent academic research has proposed a variety of open-sourced 3D mapping methods, exploring how to leverage lower-cost devices for indoor space reconstruction. These include methods based on 360-degree cameras \cite{jung2025im360texturedmeshreconstruction}, GNSS-assisted spatial localization and mapping \cite{10433646}, smartphone RGB video and LiDAR \cite{su2024rassar}, and robot-based autonomous scanning systems \cite{song2025semantic, su2024rais}. In terms of reconstruction algorithms, traditional SfM and Multi-View Stereo methods (\textit{e.g.}, \textit{COLMAP} \cite{schoenberger2016mvs, schoenberger2016sfm}) remain widely used, while recent advances in \textit{3D Gaussian Splatting} \cite{Hou2024VGaussianDM, kerbl20233d} offer new possibilities for high-quality RGB-only 3D reconstruction.

Nevertheless, most existing approaches still rely heavily on manual operation or ground-based equipment, which limits their automation capability, mapping efficiency, and scalability in large and complex indoor environments. In this work, we aim to propose an indoor mapping method that can efficiently scale to large indoor spaces with affordable hardware.

\subsection{Drone-based Indoor Mapping}
Past research \cite{karam2022microdrone,zhou2014embedded,alexovivc20233d,karam2022micro,malhotra2022fixed, 10885580,zhang2024icon} has explored the feasibility of using drones to map indoor spaces. A common research methodology in this thread is equipping drones with rich sensing capabilities or attaching additional sensors, such as RGBD cameras (\textit{e.g.,} \textit{Intel RealSense D435i}) or even LiDAR scanners (\textit{e.g.,} \textit{DJI Zenmuse L2}), to enhance spatial perception and reconstruction accuracy. In addition to academic research, commercial solutions like \textit{Skydio} \cite{skydio_3d_indoor_capture} and \textit{Flyability Elios} \cite{elios3_flyability} adopt a similar hardware strategy, leveraging LiDAR and IMU sensors mounted on drones to enable accurate indoor 3D reconstruction. While these methods and products achieve high-quality mapping results, they usually rely on expensive hardware setups and involve substantial technical effort in data acquisition and processing. Such requirements limit their scalability and general applicability, especially in scenarios where low-cost, lightweight, and easy-to-deploy solutions are desired.

In contrast to existing drone-based indoor mapping approaches, our work targets broader applicability by minimizing hardware constraints. Specifically, we propose a drone-based 3D indoor mapping pipeline that relies solely on RGB data, enabling most off-the-shelf consumer drones to be utilized for large-scale indoor reconstruction. \richard{This choice is deliberate: by demonstrating robust performance with RGB-only input, our method offers a low-cost, hardware-agnostic solution that remains practical in low-resource scenarios where high-precision stereo or depth sensors may not be available. This design lowers the barrier to adoption and extends the applicability of our approach across research, education, and industry contexts.}

\subsection{Detecting and Locating Real-world Objects}Accurately detecting and localizing key objects and facilities is essential for creating semantically rich indoor 3D maps \cite{wang2025uncertainty,lazarow2024cubify,shu2024hierarchical,zhang2024prompt3d,kolodiazhnyi2024unidet3d}. 
Traditional approaches have relied on manual annotation through GIS or digital twin tools, but these methods typically demand professional expertise and present steep learning curves for users~\cite{zhu2025knowwheregraph,rhind1988gis,ali2024enabling}. While effective in certain cases, these manual methods are time-consuming, expensive, and difficult to scale.

Recent advances in computer vision have enabled more automated approaches for object detection and localization \cite{lazarow2024cubify,shen2024cn,zhang2024prompt3d,kolodiazhnyi2024unidet3d, su2024demo}. For example, 
\textit{Project Sidewalk}~\cite{liu2024towards,weld2019deep,saha2019project} demonstrates this by crowdsourcing sidewalk feature annotations to train object detection models that automatically extract data from street views. 
Similarly, in indoor environments, \textit{RASSAR} ~\cite{su2024rassar} employs object detection with raycasting to identify and locate smaller indoor objects relevant to accessibility and safety. \richard{Among such automated approaches, \textit{YOLO (You Only Look Once)} \cite{yolov8} and its variants have become widely adopted for their efficiency and accuracy in detecting predefined object categories.}

However, existing automated approaches \cite{shen2024cn,kolodiazhnyi2024unidet3d,xu2024mvsdet} face critical limitations when applied to large and diverse indoor environments, where important objects are visually distinctive and universally defined \richard{— assumptions that often do not hold in practice. YOLO, for example, is a closed-set detector that can only recognize object categories included in its training vocabulary, limiting its applicability to customized and context-specific indoor environments.} In practice, indoor spaces often contain diverse, customized, and context-specific objects, mainly depends on the situated knowledge of local users that purely AI-driven approaches cannot easily infer or replicate.

To address these challenges, we \richard{leverage \textit{SAM2} \cite{ravi2024sam2} to operationalize} a paradigm shift in semantic indoor mapping — from treating it solely as an automated perception task to framing it as a collaborative knowledge construction process between humans and AI. Inspired by recent works on the human-robot collaboration approach for 3D semantic labeling \cite{rozenberszki20213d,chen2024synergai,zeng2024multi}, we emphasize the value of integrating human expertise with AI capabilities to create richer and more accurate indoor semantic maps.

To operationalize this vision, we introduce a human-AI collaborative annotation workflow that engages local occupants and building managers in the map creation process. Rather than relying on fixed-category, pre-trained object detectors, our approach empowers users to flexibly define and annotate objects based on their situated knowledge and contextual needs.
Ultimately, this collaborative approach repositions indoor semantic mapping as a socio-technical process — balancing the efficiency of AI automation with the interpretive agency of human users — to ensure that the resulting maps are both semantically meaningful and grounded in the complexities of real-world indoor environments.




\section{The 3D Mapping of Indoor Spaces}

FlyMeThrough is a multi-stage system that transforms RGB drone footage into interactive 3D indoor maps annotated with key Points of Interest (POIs) (\autoref{fig:teaser}). The system begins with drone-based RGB video collection, followed by Structure-from-Motion (SfM) to estimate camera trajectories and reconstruct the space into a 3D mesh. Users such as building managers then annotate key indoor facilities (\textit{e.g.}, stairs, doors, elevators) on selected video frames, informed by their grounded knowledge of the space. These annotations are expanded into per-frame video segmentation masks using SAM2 \cite{ravi2024sam2}, and these masks are subsequently projected into the 3D space using a depth-guided raycasting algorithm. The result is a semantically enriched 3D map with interactively viewable POI bounding boxes that indicate both the location and dimensions of each annotated facility. 

This work extends our previous poster paper \richard{ \cite{su2024demo}}, which explored the feasibility of using commodity drones for indoor 3D mapping and automatic facility detection of stairs, doors, elevators, and ramps. While retaining core elements such as RGB video capture and SfM-based reconstruction, FlyMeThrough introduces a more robust human-AI collaborative architecture. Our new annotation pipeline and depth-based localization significantly improve the semantic fidelity of the 3D maps, enabling end-users to meaningfully embed spatial knowledge into the resulting reconstruction.


\begin{figure}
    \centering
    \includegraphics[width=1\linewidth]{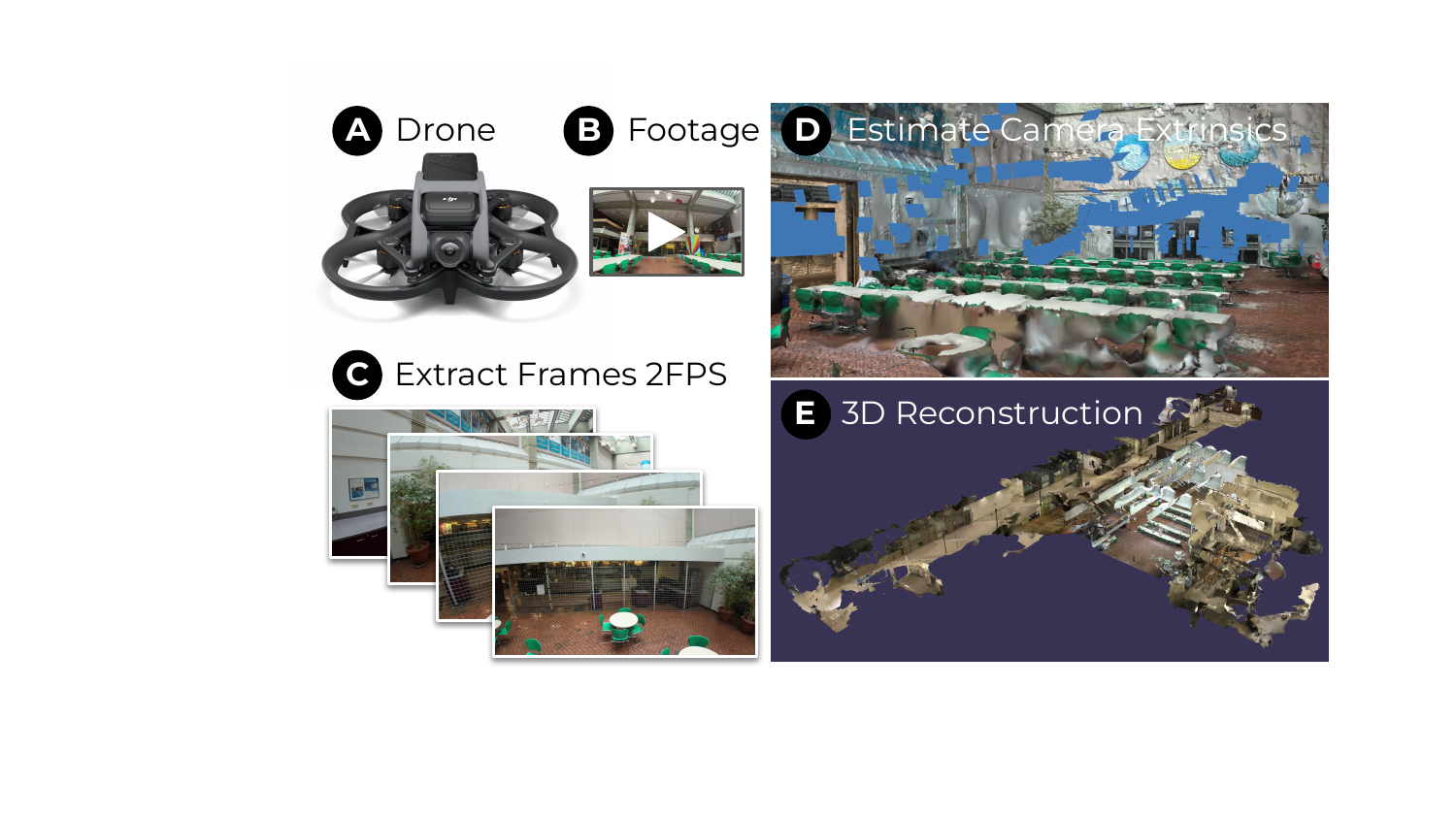}
    \caption{Drone-based indoor scanning and reconstruction. (A) The \textit{DJI Avata} drone. (B) We collect footage of flights as 4K 30FPS videos. (C) We then extract frames from the footage at 2FPS. (D) We use \textit{Agisoft Metashape} to estimate camera extrainsics of the extracted frames and also conduct (E) 3D reconstruction of the space.}
    \label{fig:reconstruction}
    \Description{A: a DJI Avata drone; B: A video footage; C. four video frames stacked; D: a inside look of a 3D reconstruction of a hall space, where blue squares indicating estimated image frame positions are up in the air, linked with black line; E: a 3D reconstruction of this space, as a T-shaped multifloor space.}
\end{figure}

\subsection{Drone-based data collection}
In line with prior work, we opt for RGB-only video input for its practicality and broad compatibility with consumer drone hardware. For safety and controllability during indoor flights, we \richard{primarily used} a \textit{DJI Avata} drone (cost \$759 when purchased in 2024) as the main testing drone for its propeller guards and immersive controller headset. We select DJI Avata's wide-angle camera setting which yields a 110-degree field of view, and capture videos during indoor flights at 4k resolution and 30 frames per second. 
\richard{To evaluate the minimum requirements for effective mapping, we also tested our pipeline with a less advanced consumer drone, the \textit{DJI Mini 2} (released in 2020, \$450), which records 2.7K, 30FPS videos with a smaller 83-degree FOV and a 12MP camera. Despite its lower specifications, the pipeline still performed effectively. Based on these tests, we recommend drones with at least a 12MP camera and 30FPS video capability for robust results.} The lead author manually controls the drone to steadily fly through the public open spaces in our tested buildings.

\subsection{SfM Indoor Reconstruction}
Following the same settings as our previous work \cite{su2024demo}, we transform the drone footage into image frames at 2 frames per second to conduct indoor reconstruction. \richard{We systematically evaluated a range of reconstruction methods, including state-of-the-art open-source solutions such as \textit{NeRF} \cite{mildenhall2021nerf,tancik2023nerfstudio}, \textit{Gaussian Splatting} \cite{kerbl20233d}, \textit{COLMAP-based SfM} \cite{schonberger2016structure, schonberger2016pixelwise}, as well as learning-based approaches like \textit{MV-DUSt3R+} \cite{tang2024mv} and \textit{MAST3R-SLAM} \cite{murai2024_mast3rslam}. }

\richard{Our experiments revealed that COLMAP struggles to reconstruct building-scale indoor scenes due to cumulative drift and the lack of loop closure \cite{schonberger2016structure}, which leads to error propagation in large and repetitive environments. NeRF \cite{mildenhall2021nerf} and Gaussian Splatting \cite{kerbl20233d}, which rely on COLMAP for camera pose estimation, inherit these limitations and exhibit degraded performance in large-scale settings. SLAM systems such as \textit{DROID-SLAM} \cite{teed2021droid} reconstruct large-scale scenes online and refine them through global optimization, but their accuracy remains constrained by the quality of initial estimates. Recent feedforward approaches like MAST3R-SLAM \cite{murai2024_mast3rslam}, which leverage learned priors for reconstruction, operate on two-frame inputs and often produce inconsistent geometry that is difficult to correct in post-hoc optimization.}

\richard{Based on these evaluations, we selected the proprietary photogrammetry software Agisoft MetaShape \cite{over2021processing} as the most robust option in terms of output quality, reconstruction success rate, and computational efficiency. MetaShape performs SfM camera pose estimation and 3D reconstruction reliably for large and complex indoor spaces. Figure \ref{fig:reconstruction}E shows the reconstructed 3D model, and Figure \ref{fig:reconstruction}D shows the estimated camera pose and flight trajectory.}

\subsection{Indoor Facility Segmentation}
To embed key facility information into the reconstructed 3D map, FlyMeThrough employs the SAM2 \cite{ravi2024sam2} video segmentation model to interactively identify and label key indoor objects. For each scene, users make intuitive clicks on the video frames to annotate key indoor objects, including pre-selected types drawn from existing indoor scanning work \cite{su2024rassar}: \textit{door}, \textit{elevator}, \textit{ramp}, \textit{stairs},\textit{etc.} , as well as any customized categories named by users. 

\begin{figure}
    \centering
    \includegraphics[width=1\linewidth]{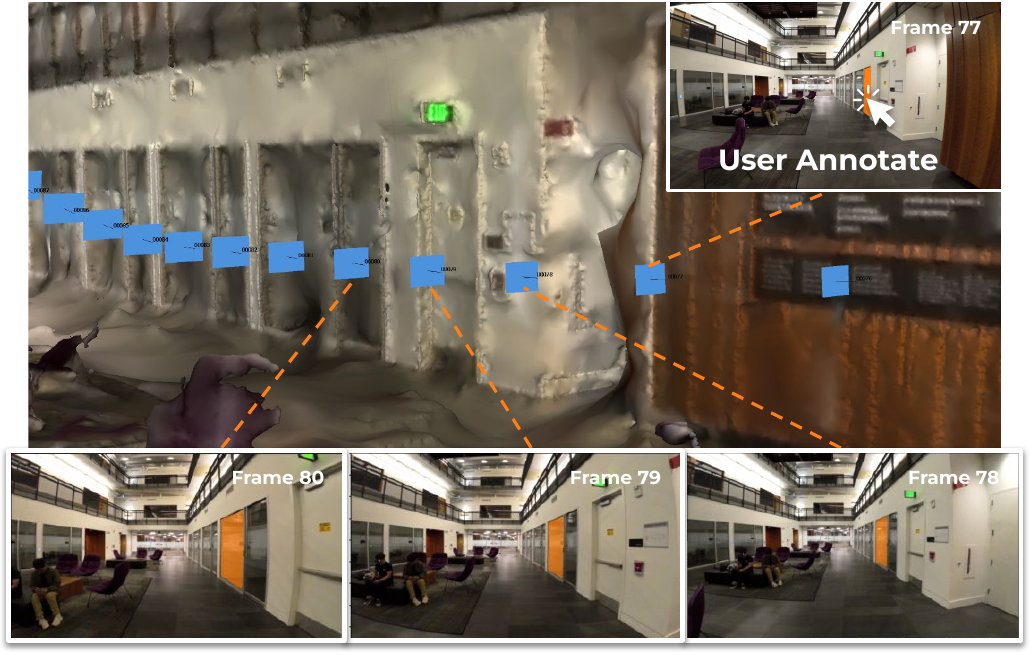}
    \caption{We employ SAM2 to segment key indoor facilities. When user annotate in one frame, the mask will be propagated to subsequent frames.}
    \label{fig:segmentation}
    \Description{An image showing a indoor reconstruction scene with multiple blue image frames in the middle. One frame which says "User Annotate" shows a user annotation action, where a mouse click icon is upon an orange region indicating the selected mask. There are three other enlarged frames on the bottom, showing the same scene but in slightly different viewing position, each showing the annotated part in orange, but in slightly different size.}
\end{figure}

This design marks a departure from our previous system  \cite{su2024demo}, which relied on YOLOv8 \cite{yolov8} for fully automated POI detection. While effective in automating object recognition, YOLOv8 lacked flexibility for open-vocabulary object types and precision for multi-frame consistency, especially in novel or complex indoor scenes. In contrast, SAM2 offers several key advantages that informed our decision to adopt it:

\textbf{Interactive and Efficient Annotation}: SAM2 natively supports interactive prompts—such as point clicks and bounding boxes — which integrate seamlessly into our UI and provide intuitive user interaction while minimizing annotation effort. With minimal user input, the model generates high-quality pixel-level masks, significantly reducing manual efforts.

\textbf{Generalization and Adaptability}: While YOLOv8 is limited to closed-set object categories, SAM2 shows strong generalization across diverse object types and indoor layouts. Its zero-shot segmentation capability enables accurate labeling of novel object instances—ideal for dynamic and unstructured indoor settings as well as varied user requirements.

\textbf{Precision and Continuity}: SAM2 produces pixel-accurate segmentations that capture fine object boundaries. This granularity is critical for downstream ray-casting-based localization, where geometric precision determines placement accuracy in the 3D space. Also, the results are video segmentations that propagate across adjacent frames, which improve temporal consistency for the following localization step. 

In deployment, we employ a pretrained SAM2 checkpoint~\cite{sam2_github}, which tokenizes all input video frames and infers object masks in following frames based on a user-provided segmentation (See \autoref{fig:segmentation} for an example). To enhance temporal consistency, we adopt a local temporal control strategy. Specifically, segmentation predictions of a user annotation are retained only for consecutive frames and are terminated when the target object exits the view. This helps mitigate missegmentations caused by repetitive layouts—a common challenge in hallways or mirrored environments.

\subsection{Localizing Indoor Features with Depth-guided Ray-casting}
\label{subsec: raycasting}
The segmentation results, as masks over sequential video frames, are projected into the reconstructed 3D maps for localization. In our previous system, this projection was performed by individual pixels following the classical pinhole camera model \cite{sturm2021pinhole}. Each masked pixel was back-projected through the camera’s intrinsic and extrinsic parameters to cast a ray into the 3D scene. Intersections between these rays and the reconstructed surface mesh were computed, and the resulting hit points were post-processed via spatial clustering to generate 3D bounding boxes for each object instance. However, this approach was highly sensitive to mesh artifacts, such as surface holes and noise, which frequently led to fragmented or inaccurate localization results. 

To address this, we introduce a \textbf{depth-guided ray-casting strategy} that leverages monocular depth estimation to constrain ray lengths within each image mask. This process generates a segmentation point cloud to be raycasted as a whole, thereby significantly improving localization accuracy and suppressing false intersections caused by mesh artifacts. 

\begin{figure}
    \centering
    \includegraphics[width=1\linewidth]{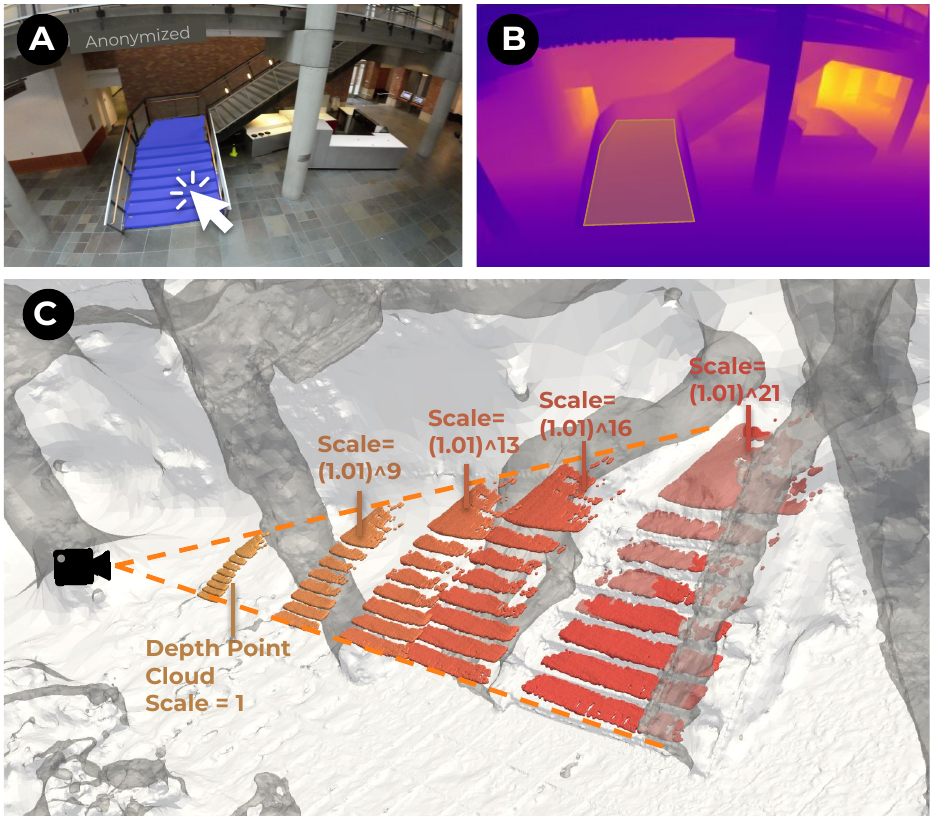}
    \caption{We employ a depth-guided ray-casting to locate user annotations. (A) Each annotation mask will be incorporated with (B) inferred depth map to produce a 3D point cloud, (C) which gets iteratively enlarged until intersection with the reconstructed 3D model.}
    \label{fig:casting}
    \Description{Top row shows an annotating scene where a mouse icon click on a stair and get the stair highlighted in purple. Then the mask gets applies to a depth map. Bottom image shows a casting process that turns this masked depth information into a pointcloud and iteratively enlarge it to intersect with reconstruction mesh.}
\end{figure}

\textbf{Classical Pinhole Projection Model} We retain the pinhole camera model, where a 3D point \( \vec{P}_w = (X_w, Y_w, Z_w)^T \) is projected to a image coordinate \( (u, v) \) with camera intrinsic matrix \( K \) and extrinsic parameters \( R \) and \( t \), following:

\begin{equation}
Z_c \cdot 
\begin{bmatrix}
u \\
v \\
1
\end{bmatrix}
=
K \cdot 
\left[ R \,\middle|\, t \right]
\cdot
\begin{bmatrix}
X_w \\
Y_w \\
Z_w \\
1
\end{bmatrix}
\tag{1}
\end{equation}

Given known \( K \), \( R \), and \( t \) from the SfM reconstruction, the only missing component is the depth \( Z_c \) value, which is traditionally calculated by raycasting. 

\textbf{Create Segmentation Point Cloud with Depth Estimation} Although depth is not directly available in monocular RGB videos, we can still estimate a relative depth for each pixel point in the same image frame using a state-of-the-art monocular depth prediction model Depth Pro~\cite{bochkovskii2024depth}. We then use the predicted depth values, indicated as \( \hat{Z}_c \), to extrude the pixels in the segmentation mask to 3D coordinates \(\tilde{\vec{P}}_w\):

\begin{equation}
\tilde{\vec{P}}_w = R^{-1} \cdot  
\left(
\hat{Z}_c \cdot K^{-1}
\begin{bmatrix}
u \\
v \\
1
\end{bmatrix}
- t
\right)
\tag{2}
\end{equation}

This process produces a 3D point cloud from pixels in a 2D segmentation mask, which captures the spatial extent and relative depth landscape of the object in the scene. See \autoref{fig:casting} for an example. Note that the resulting point cloud is not metrically scaled with the mesh model, as it is derived from predicted monocular depth with an unknown global scale factor, which we iteratively approximate in the following casting steps.

\textbf{Unified Segmentation Point Cloud Casting} Rather than raycasting each individual pixel (as in our prior approach) which can be stopped by irregular geometry or go through holes, we cast the segmentation point cloud as a whole. See \autoref{fig:casting}. We iteratively exponentiate the scale of the point cloud by 1.01 and calculate the correlation portion with the reconstructed mesh, until it reaches a threshold (currently set as a fixed 22\% based on experiments).
This casting strategy avoids sensitivity to local mesh defects (holes, clutter, irregular geometry), and helps preserve the object's shape and orientation in the casting results. This prevents premature or delayed intersection errors, and mitigates offset artifacts introduced by surface irregularities. Compared with our previous system, the casting process achieves higher geometric accuracy and better semantic consistency within complex indoor 3D models. 

\textbf{Adaptive Downsampling for Efficiency}
To optimize runtime performance during the cloud-based raycasting steps, we introduce adaptive spatial downsampling. We reduce the point cloud size while preserving its overall shape: The number of retained points is dynamically selected. Larger objects retain more points to ensure precision, while smaller ones are compactly represented. This balance significantly reduces computation while maintaining sufficient spatial fidelity.

\subsection{Bounding Box creation}

After raycasting the downsampled segmentation cloud into the mesh, the resulting intersection points are clustered to generate a final 3D bounding box. This bounding box encapsulates not only the object's location, but also its approximate size and orientation in 3D space.

Since we raycast video segmentation results across multiple frames for the same object to further even out noise, this multi-frame casting result is a densely overlapping point cloud data. We apply \textit{DBSCAN} ~\cite{ester1996density} to segment and filter the point cloud, removing outliers and preserving the most coherent geometric regions associated with the object. We then compute a PCA-based approximation of the Minimum Volume Bounding Box ~\cite{dimitrov2006bounding} for each cluster, which produces tighter, direction-aware bounding volumes that better reflect the true geometric extent of the object, especially under non-uniform orientations.


\subsection{Modularity of Implementation}
As a multi-step pipeline that includes commodity hardware, commercial software, classical algorithms, state-of-the-art models, as well as our original algorithms and methods, we realize the potential limitation and space for future improvements in all these components. In this case, we have designed our system to be modularized so that each step or component can be replaced with better-performing versions in the future. \richard{For example, while our current implementation uses monocular RGB inputs with estimated depth, the depth-estimation module can be seamlessly swapped with real depth data if the drone is equipped with onboard depth sensors. Similarly, the 3D reconstruction module—currently based on commercial photogrammetry software—can also be replaced with open-source or more advanced reconstruction methods as they become available.} This modularized architecture enables customized implementation by future users to better adapt to their specific hardware, software, and also compute power limitations. 


\section{The FlyMeThrough Interface}

\begin{figure*}
    \centering
    \includegraphics[width=1\linewidth]{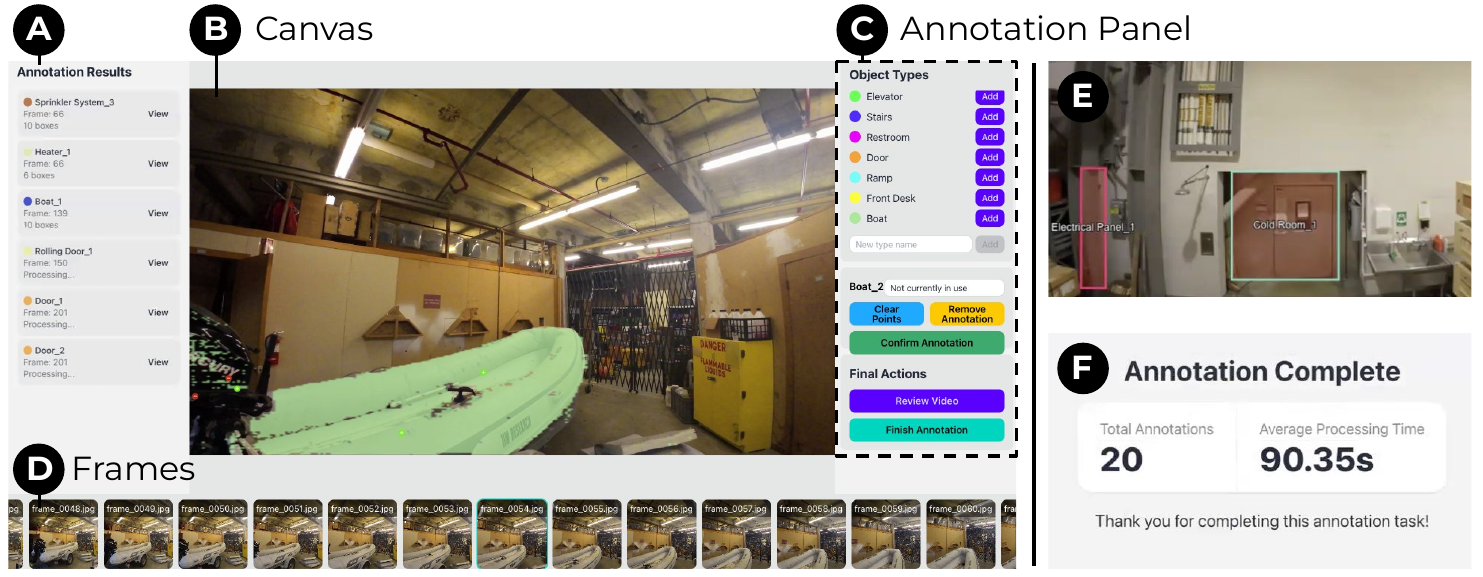}
    \caption{The annotation interface. (A) Annotation results panel shows all confirmed annotations. (B) Canvas shows the selected frames and enables click-based annotation for segmentation masks, which will be highlighted with color. (C) Annotation panel shows object types to choose from and annotation actions like \textit{Clear Points, Comfirm Annotation}, as well as final actions like \textit{Finish Annotation}. (D) Frames panel lists all video frames available for annotation. (E) After user confirm an annotation, it gets processed by our segmentation server in real time and the returned segmentation results for subsequent frames will be shown as bounding boxes. (F) After finishing the annotation, interface shows a summary.}
    \label{fig:annotation}
    \Description{Image shows a UI screenshot. The middle part shows a warehouse scene with a inflated boat being highlighted with green. Left part lists several finished annotations. Right part shows the annotation panel, which include some pre-listed object types, as well as input box for user to create new types. E shows an image with two bounding boxes, and F shows a summary of the annotation, saying "Total Annotations 20, Average Processing Time 90.35s".}
\end{figure*}

As elaborated in the previous section, FlyMeThrough engages human users to provide ground knowledge, \textit{i.e.,} identifying what and where the most important and relevant indoor facilities are in the reconstructed indoor 3D maps. This crucial mapping task cannot be fully automated with computer vision due to two reasons. For one, facilities in indoor spaces vary in appearance, making them hard to be accurately detected by generalized models; For another, the functions and importance of indoor facilities are highly specific and personalized in the site-specific practices. In this case, we designed and implemented an annotation interface to enable human-AI collaborative creation of POIs in reconstructed 3D indoor maps. We also created a review interface that enables users to review the output maps as interactive 3D models.

\subsection{Annotation interface}
We create a web interface that enables efficient and intuitive indoor POI annotation. Unlike traditional indoor mapping process that leverages CAD files and requires relevant design skills, our web interface shows the image frames of the drone footages and enables click-based annotation operations to enable user to mark any target objects, like entrances, doors, and stairs, in the image frames. \autoref{fig:annotation} shows the annotation interface, which include four main components: Annotation canvas (\autoref{fig:annotation}B) that shows image frames, annotation points and masks, as well as bounding boxes of processing results of annotations. Frames panel  (\autoref{fig:annotation}D), which lists image frames of the drone footage in the order of time. Annotation tools panel  (\autoref{fig:annotation}C), which includes types of objects for users to select from, and action buttons for specific annotations and the overall annotation task. Annotation results panel (\autoref{fig:annotation}A), which lists the user-confirmed annotations, their processing progress, and also enables reviewing of annotations.

When annotating, users browse through the frames panel and select video frames to annotate. They can create object instances with the annotation tools panel from either pre-defined object types, or by creating new object types. For each object instance, users can optionally input a more detailed description and click on the annotation canvas to select the target object. A SAM2 \cite{ravi2024sam2} model embedded in the web interface will provide a real-time segmentation mask, which can be revised by adding more positive points or negative points. Positive points are added by clicking on the non-masked parts of the canvas, and negative points by clicking on the masked parts. Users can also clear points to restart, remove the instance entirely, or, when they are satisfied with the annotation, click confirm to send this annotation to our server, where the mask will be propagated to subsequent video frames by video segmentation \cite{ravi2024sam2}. The segmentation results, when finished, will be visualized as bounding boxes in the subsequent frames (\autoref{fig:annotation}E), and also further processed by raycasting (see \autoref{subsec: raycasting}) to create 3D bounding boxes in the reconstructed 3D mesh. 

\begin{figure}
    \centering
    \includegraphics[width=1\linewidth]{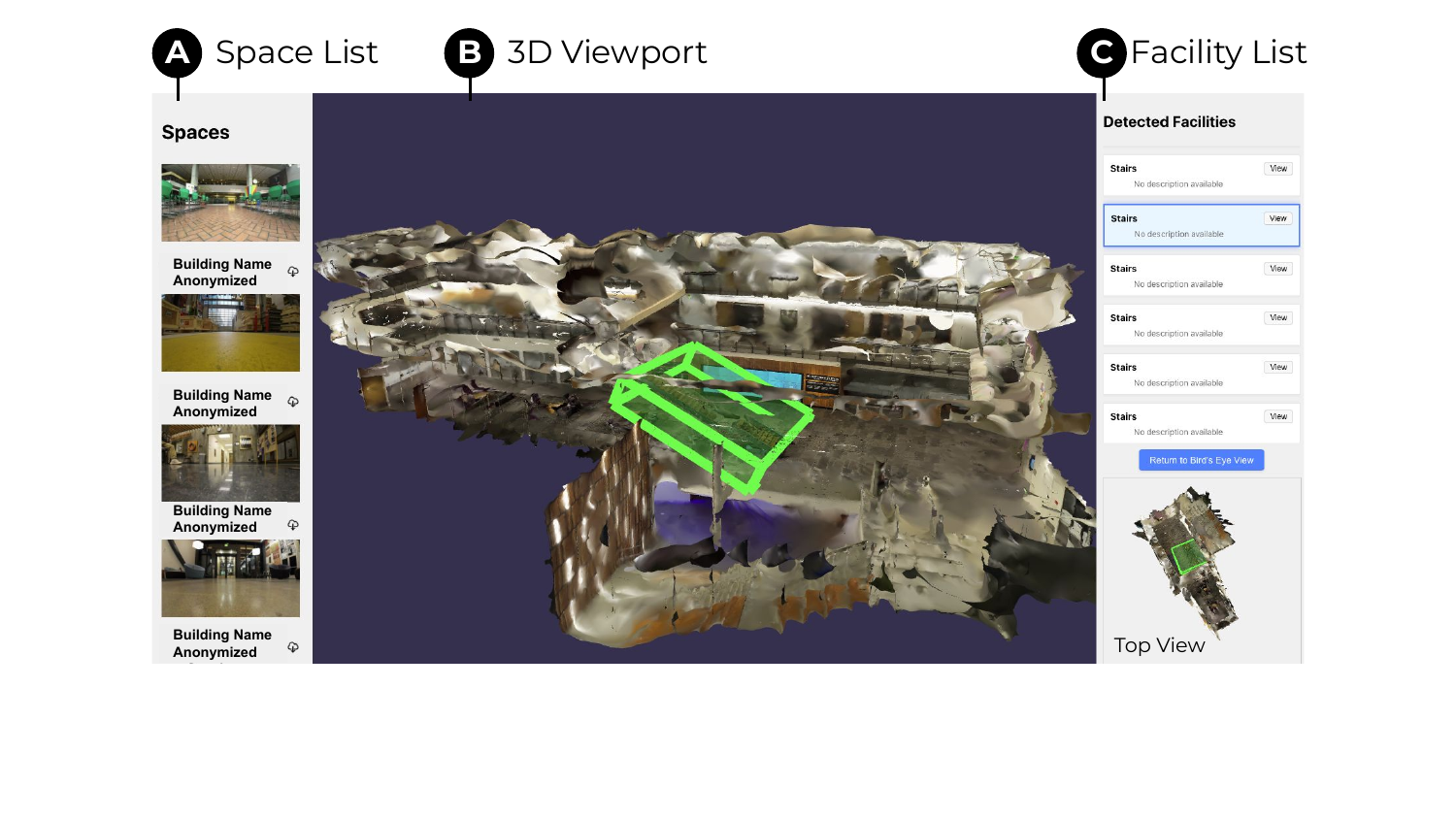}
    \caption{The review interface. (A) A list of spaces available to review. (B) A 3D viewport shows the 3D map and annotated objects. (C) List of annotated facilities, each can be highlighted. Click on the \textit{"View"} button will focus the view to the bounding box. Interface also provide a button to go back to bird view, and an orthographic top view of the map as a floor plan of the space.}
    \label{fig:review}
    \Description{A reviewing interface with a list of spaces on the left, a 3D viewport in the middle, showing a multi-floor space with a green bounding box being highlighted in the middle. Right shows a facility list, with the "Stair" object being selected.}
\end{figure}

\subsection{Visualization interface}
We also implement another web interface to visualize the 3D indoor maps, which include 3D mesh models that reconstruct the scanned indoor space, and also 3D bounding boxes that mark the location and dimension of annotated indoor objects. \autoref{fig:review} shows this interface, containing a space list panel (\autoref{fig:review}A) for available indoor maps, a 3D interactive canvas (\autoref{fig:review}B) that shows the indoor map, and also a right panel (\autoref{fig:review}C) that lists all annotated facilities, as well as a floor plan view of the 3D map rendered as a top-down orthogonal view of the 3D map. Users can click and drag to rotate view angles, browse through a list of objects, hover on them to see them highlighted in the 3D canvas, and also click any object to focus the view on it.

\section{Evaluation}
We evaluate FlyMeThrough for its technical performance and overall usability. We collected drone footage of 12 spaces with varying space types, sizes, vertical heights, and also functionalities. We process these spaces into 3D indoor maps to understand the technical capability of our pipeline. We also evaluate interface usability with a study among 10 participants, including five building managers and five building occupants, which also aims to understand the practical implications and potential barriers to implementing drone-based indoor mapping systems.

\subsection{Procedure}
In order to conduct indoor drone flights, we first reached out to building managers to authorize and supervise us to fly our drone in their buildings. With approval, the lead author manually controlled the drone to capture building spaces at off-hours to minimize risk and disruption. We collected flight footage of 12 indoor spaces with varying sizes, space types, and building functionalities. See \autoref{tab:space_summary} for more details. We then reconstructed the spaces with our system to test robustness. 

\begin{figure*}
    \centering
    \includegraphics[width=1\linewidth]{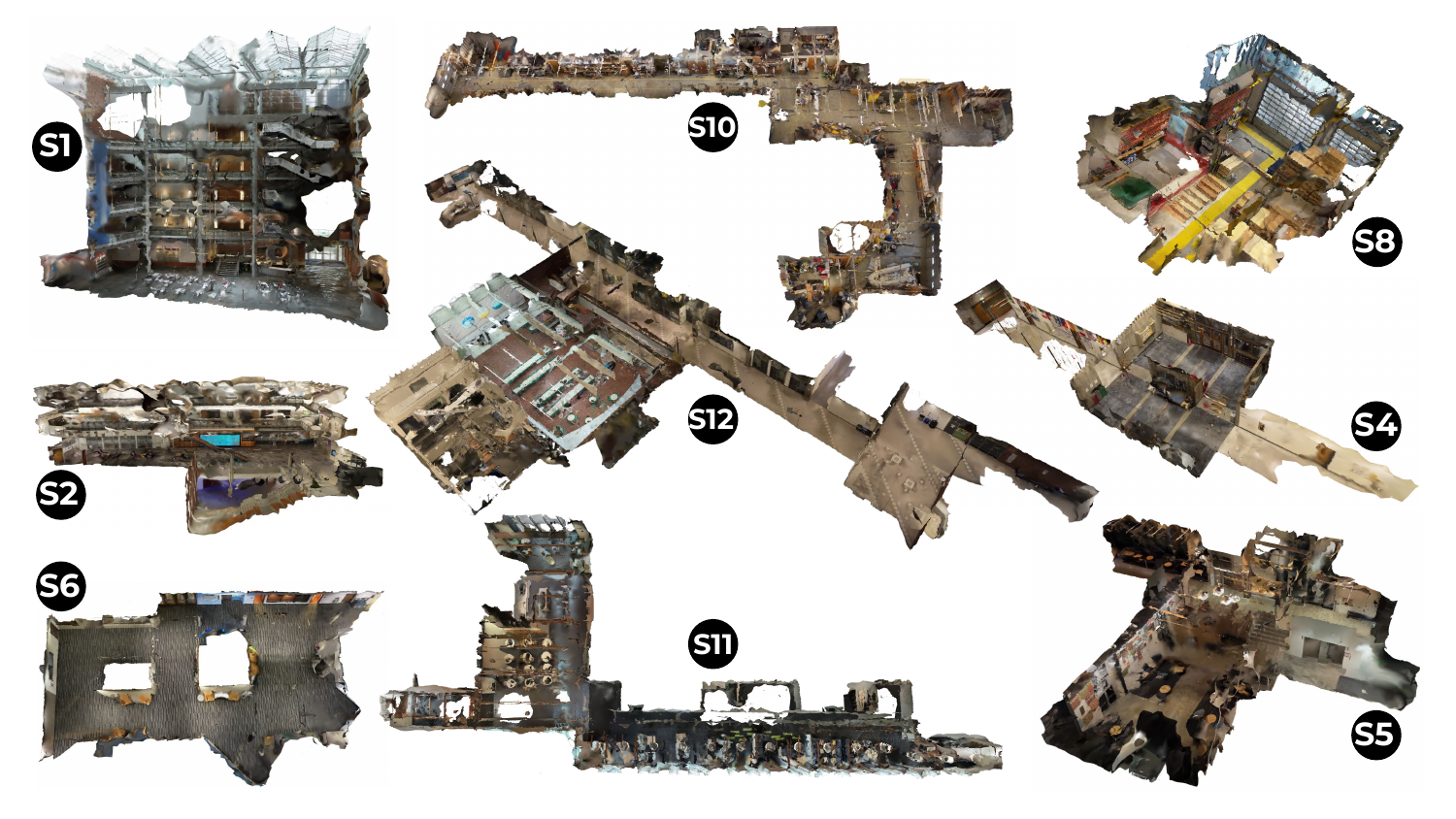}
    \caption{Nine of the scanned spaces.}
    \label{fig:spaces}
    \Description{A gridview of 9 reconstructed spaces, each shows the space in a top-down or side view.}
\end{figure*}

\begin{table*}[ht]
\centering
\caption{Overview of Spaces with Video Data and Reconstruction Status}
\label{tab:space_summary}
\begin{tabular}{cccccc}
\toprule
\textbf{Space ID} & \textbf{Function} & \textbf{Area (sqm)} & \textbf{Height (floors)} & \textbf{Video Length} & \textbf{Reconstruction Status} \\
\midrule
\rowcolor[HTML]{EFEFEF} 
S1  & Education   & 1600  & 6 & 3min45s & Good   \\
S2  & Education   & 1400  & 4 & 2min33s & Good   \\
\rowcolor[HTML]{EFEFEF} 
S3  & Education   & 100   & 1 & 2min08s & Good   \\
S4  & Office      & 120   & 1 & 3min22s & Good   \\
\rowcolor[HTML]{EFEFEF} 
S5  & Education   & 180   & 1 & 2min10s & Good   \\
S6  & Education   & 600   & 1 & 3min06s & Good   \\
\rowcolor[HTML]{EFEFEF} 
S7  & Exhibition  & 3000  & 2 & 6min07s & Good   \\
S8  & Engineering & 260   & 3 & 5min03s & Good   \\
\rowcolor[HTML]{EFEFEF} 
S9  & Office      & 1100  & 4 & 5min33s & Failed \\
S10 & Engineering & 2000  & 1 & 5min53s & Good   \\
\rowcolor[HTML]{EFEFEF} 
S11 & Education   & 800   & 1 & 4min02s & Good   \\
S12 & Education   & 2400  & 3 & 9min13s & Good   \\
\bottomrule
\end{tabular}
\end{table*}

To evaluate the usability and performance of our interface, we conduct a user study which guides participants to conduct POI annotation and 3D map review. The study is conducted through both remote and in-person interviews, consisting of five sequential parts: (1) initial set up, (2) drone mapping demonstration, (3) annotation interface experience, (4) initial interview and discussion, and (5) final model review and feedback. Each session lasted \textasciitilde60 minutes.

The first part begins with meeting the participant, reviewing and obtaining informed consent, and conducting a brief background survey. Building managers are asked additional questions about their current building evaluation
practices.
In the second part, if the participants have not witnessed the drone flying process, they are shown a comprehensive demonstration of the drone-based mapping process through recorded footage from both the drone's perspective and a bystander's view. 
This dual-view presentation help participants understand both the drone's operational capabilities and its presence in the space.
The third part focused on the annotation interface, where participants were given hands-on experience with the annotation process. 
In the fourth part, we ask a set of open-ended questions to assess participants' reactions to the drone flight process, evaluate the annotation system's usability, and explore how this technology could be integrated into their existing building evaluation practices or provide value in other applications.
During this discussion period, the system processes the collected data in the second part into an instance-embedded 3D model.
The final part began once the 3D model is ready. 
We demonstrate the final result through our interactive interface, showing participants how the drone-captured data translates into a usable building model. 
The session concluds with additional questions about their overall experience with the system and their assessment of its performance in relation to their respective needs.

\subsection{Participants}
We recruited five building managers to support our drone flights. All of these building managers are also interviewed in our user study. Additionally, we recruited five building occupants who are active users of the scanned buildings. Demographics. Among these participants, four witnessed our drone operations, and the rest watched video recordings. Participants are compensated \$25 per hour for guiding and witnessing the drone flights and for participating in our user study. See \autoref{tab:demographics} for more demographics information.

\begin{table*}[ht]
\centering
\caption{Participant Demographics and Annotation Performance}
\label{tab:demographics}
\resizebox{\textwidth}{!}{%
\begin{tabular}{ccccccccc}
\toprule
\textbf{Participant} & \textbf{Age} & \textbf{Gender} & \textbf{Role}           & \textbf{Space Tested} & \textbf{Experience (yrs)} & \textbf{Footage/Witness} & \textbf{Annotations} & \textbf{Success (\%)} \\
\midrule
\rowcolor[HTML]{EFEFEF} 
M1   & 35-44 & M & Building Manager  & S1      & 13   & Footage & 12 & 100.00 \\
M2   & 45-54 & M & Building Manager  & S8, S10 & 7    & Witness & 33 & 75.76  \\
\rowcolor[HTML]{EFEFEF} 
M3   & 25-34 & M & Building Manager  & S2      & 9    & Footage & 19 & 100.00 \\
M4   & 18-24 & F & Building Manager  & S11     & 1.5  & Witness & 16 & 87.50  \\
\rowcolor[HTML]{EFEFEF} 
M5  & 55+   & M & Building Manager  & S12     & 7    & Footage & 9  & 100.00 \\
O1   & 25-34 & M & Occupant              & S6      & 5    & Witness & 29 & 93.10  \\
\rowcolor[HTML]{EFEFEF} 
O2   & 25-34 & M & Occupant              & S2      & 5    & Footage & 8  & 87.50  \\
O3   & 25-34 & M & Occupant              & S4      & 3.5  & Witness & 8  & 100.00 \\
\rowcolor[HTML]{EFEFEF} 
O4   & 25-34 & M & Occupant              & S5      & 0.5  & Footage & 14 & 100.00 \\
O5   & 25-34 & F & Occupant              & S1      & 6    & Footage & 10 & 100.00 \\
\bottomrule
\end{tabular}%
}
\end{table*}

\subsection{Analysis Approach}
Our analysis of the semi-structured interviews focused on summarizing high-level themes. 
One researcher developed a set of themes through qualitative open coding~\cite{charmaz2006constructing} based on the video transcript, then coded the responses according to the themes. 
Participant quotes have been slightly modified for concision, grammar, and anonymity.

All annotations, segmentations, as well as subsequent raycasting results, are logged and recorded during user studies. The research team analyzes such data for computation efficiency and robustness.

\subsection{Technical Results}
\begin{figure}
    \centering
    \includegraphics[width=1\linewidth]{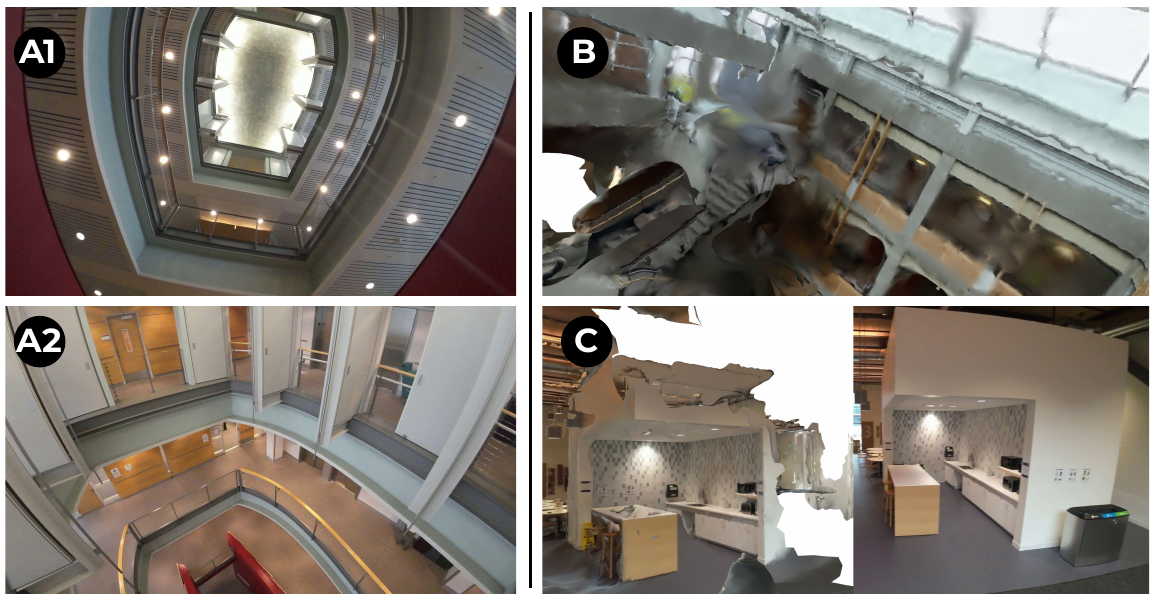}
    \caption{Three types of failure cases. (A) A1 and A2 show two images from S9, which is an office building with a multi-floor atrium space. The reconstruction of this space failed. (B) The reconstructed 3D map can be blurry on less scanned parts. (C) Left shows the reconstruction missing certain parts, \textit{e.g.} the trash collector shown in the original footage image shown on the right.}
    \label{fig:failure}
    \Description{Left side shows two indoor images of a boat-shaped atrium space, where floor structures are repetitive. Right top (B) shows a corner of a 3D reconstruction and it's blurry; C shows a comparison of two images, left is a partial reconstruction with white blank spaces on the right; right shows the original camera capture with a trash collector that's not seen in the left reconstruction.}
\end{figure}

Among the 12 target spaces, reconstruction succeeded in 11 cases, with only 1 failure. \richard{The single failure (S9) likely resulted from its challenging architectural characteristics: a circular, multi-floor corridor with highly repetitive vertical and horizontal structures, which may have hindered reliable feature detection and matching during the SfM process (see \autoref{tab:space_summary} for space details and \autoref{fig:failure}A for pictures).} \autoref{fig:spaces} shows nine of the 11 reconstruction results. During the user study sessions, participants created an average of 15.8 annotations per space (individual counts are provided in \autoref{tab:demographics}). All annotations were successfully processed by our SAM2 model, with an average processing time of 41.10 seconds per annotation (SD = 11.21). For the ray-casting module, each casting operation took an average of 63.6 seconds (SD = 48.65). Notably, 91.77\% of all annotations were successfully cast into bounding boxes for users to review. Annotation counts and casting success rates per participant are summarized in \autoref{tab:demographics}.

We also analyzed the failed casting cases. A total of 13 failures were recorded, of which 8 (61.5\%) were due to missing camera parameters in the corresponding frames at the beginning of the SfM-based indoor reconstruction process. 
Another five additional failures (38.5\%) were caused by SfM reconstruction being partially incomplete in some cases, leading to missing geometry. See \autoref{fig:failure} B and C for examples. 

\subsection{Findings}
A total of 158 objects were annotated by the participants. Among them, doors (including regular and rolling doors) were the most frequently labeled (31), followed by common building infrastructure such as stairs (21), entrances (10), elevators (9), restrooms (5), and ramps (3). In addition to these core architectural features, participants also annotated a wide range of smaller objects (\textit{e.g.}, signage, furniture, equipment).
We also observed a clear distinction in annotation focus between building managers and building tenants: managers tended to label more technical and safety-related features (\textit{e.g.}, fire control systems, equipment panels), whereas users more frequently annotated everyday-use elements such as sofas, whiteboards, and trash bins.

Below we summarize the major themes that emerged from our user interviews.

\subsubsection{Annotation System Usability and Areas of Improvement}

Our evaluation of FlyMeThrough's annotation interface revealed generally positive feedback regarding its usability.
When asked to rate ease of use on a 7-point Likert scale (1=very difficult, 7=very easy), participants reported a median rating of 6 (\textit{IQR}=2.00), indicating an overall favorable assessment of the system's usability.
Several participants explicitly characterized the interface as intuitive, with building managers and occupants noting it was \sayit{pretty easy to use} (M2, O5) and \sayit{not complicated at all} (M5). 
This suggests that FlyMeThrough's annotation workflow can accommodate users with varying levels of technical expertise.

Despite overall positive reception, we also identified areas of improvement. A common challenge involved object selection workflow: \sayit{The confusing thing is you have to click this first and then identify the thing again... the order is maybe a little bit confusing} (M1). 
Similarly, M4 noted: \sayit{ Sometimes I would forget to [select the object type first]. I would click on an object and then realize I didn't select what type of object it was.}
Participants also expressed interest in additional functionality that could enhance their experience. 
For example, O2 suggested implementing an example-based approach: \sayit{I think for the annotation part, it might be helpful if I can be provided with examples so I don't have to start on my own. I can just say yes or no.}
This comment points to the potential benefit of incorporating automated annotation features with human-verification.

\subsubsection{Perceived Model Quality}

When asked to rate the quality of the model on a scale of 1-7 (1=poor, 7=excellent), participants had a median rating of 5 (\textit{IQR}=0.75), indicating generally positive perceptions of the system's output quality. 
We also asked participants to give a binary rating (accurate/inaccurate) to the location and size of the bounding box for each of the annotated objects. The average percentage of objects rated as accurate was 71.54\% (\textit{SD}=29.73\%). This wide standard deviation reflects considerable variation in how participants perceived the annotation results.

Some participants found the model's output to be satisfactory and accurate for their needs. However, others expected more precise geometric alignment, particularly with architectural elements. As one building manager explained: \sayit{so you end up with a larger bounding box and it's at an angle relative to where you're viewing the individual item, as opposed to the item itself being a rectangular prism up in the ceiling} (M2). 
Similarly, M5 expressed that he was hoping the bounding box would exactly fit the volume of the room: \sayit{the box doesn't line up with this atrium exactly}.
This contrast in expectations was also contributed by the diverse scale of objects participants labeled, ranging from entire spaces like atria to small items such as electrical panels.
These findings suggest that while the current model quality is sufficient for many use cases, expectations for geometric precision vary considerably across users and object types.

\subsubsection{Concerns about Drone Flying Technology}
When asked if they had concerns about flying the drone, participants were relatively relaxed.
They primarily focused on appropriate scanning locations and privacy considerations rather than expressing significant safety worries.
Participants suggested conducting scans during off-hours and limiting them to public areas: \sayit{common spaces would be [appropriate to scan]. Like we could do the hallways, we could do the large atria} (M1). Multiple building managers (M1, M2, M3) emphasized avoiding sensitive spaces such as deans' offices, research labs, and dormitories.
Privacy emerged as a key consideration. 
M4 recommended providing advance notice to building tenants before scanning so individuals could avoid the area if they did not wish to be included. 
Building occupants raised concerns about facial privacy, with one suggesting that \sayit{maybe it's best to blur people's faces before annotation} (O5).

\subsubsection{Diverse Use Cases and Stakeholder Benefits}

Participants demonstrated diverse needs that could be fulfilled by FlyMeThrough. Building managers and occupants identified distinct use cases, with several participants recognizing potential benefits beyond their own user group.

\textbf{For Building Managers.}
Building managers identified numerous practical applications for FlyMeThrough, ranging from daily operational tasks to strategic space planning and resource management.
Our interviews revealed that FlyMeThrough has strong potential to enhance current mapping practices. 
One manager highlighted how the system could eliminate redundant site visits: \sayit{Sometimes I'll go look at a room or space and then if I don't take a picture I go back to my office. If I forget then I just have to go back. But it's nice to have all the data right here} (M1). 
This suggests that having comprehensive spatial data readily available could improve workflow efficiency.
The system's potential for remote inspection was particularly valued for maintenance operations.
M1 explained: \sayit{It just saves us from having to go check on things, for example, there's a leak from the roof}. 
Another specifically highlighted specialized applications: \sayit{This would come in most handy from a facilities perspective if we did the 3D mapping in a mechanical room, where all the pumps and valves, piping and stuff like that} (M5). 
These comments suggest that the system could reduce the need for in-person inspections of routine issues and provide valuable documentation of complex mechanical spaces.

Beyond maintenance, managers saw potential for enhancing event planning and coordination.
One manager shared: \sayit{All the images I have of the atrium are just me taking pictures with my phone from different angles. But it could be really helpful to just have this. Then for example, if I have an event coming in, I could say to them, here's the space. We could mount lights here on this column and this column} (M1).
Training applications were also identified as valuable use cases. 
As M3 noted: \sayit{If for whatever reason I cannot meet with somebody for a health and safety training, I could do it virtually with them or for them to be able to review that space after the safety training as a records} (M3). This points to potential educational applications beyond basic navigation and space management.

At a more strategic level, managers recognized the system's potential for space utilization analysis.
As M2 noted: \sayit{What's the square footage that we're using for storage versus classrooms versus offices versus labs? Those are things that are actually important to the university from a sense of when grants go in and charging rates to grants for infrastructure} (M2). This highlights FlyMeThrough's potential to support not only operational decision-making but also financial planning and resource allocation.

\textbf{For Building Occupants}
Building occupants focused primarily on navigation applications. 
One noted potential use in large, complex spaces: \sayit{For conference, if you go to convention center, it's super huge. Also for direction} (O2). The same participant highlighted accessibility benefits: \sayit{Maybe if I went to a new country that I don't speak their language, but I want to go to a museum. I want to see a specific artifacts, I want to go to a specific section. I can just use this. Not reading anything} (O2). This suggests that visual-based navigation could overcome language barriers in unfamiliar environments.

Essential amenities and safety information were prioritized by occupants: \sayit{When I go to a new building, I always want to know the emergency exit and also bathrooms and water. Those are the most important thing. And whether there are elevators or stairs to go to the other place} (O5). 
This emphasizes the importance of including basic amenities and safety features in the annotation system.

\textbf{Cross-User Benefits}
Notably, many participants recognized potential benefits beyond their immediate needs. One manager articulated how the system could serve multiple stakeholders: \sayit{It would be the kind of space that would actually work really well for people at all level of the department. Grad students could see their office before they move in. Facilities could get a chance to look at a space without having to spend 30 minutes getting a car and driving across campus. It could work in group meetings talking about space allocation, you could pull this up and have a conversation in a meeting space as opposed to having to walk through it. So everybody that uses the space could find a way to benefit from it} (M2).
This perspective suggests that FlyMeThrough has the potential to serve as a unified spatial information platform that bridges the needs of different stakeholder groups, from administrators and facilities staff to everyday users and visitors.

\section{Discussion \& Future Work}
In this paper, we introduce \textit{FlyMeThrough}, a drone-based indoor mapping system that leverages commodity drones to scan indoor spaces and creates interactive and POI-infused 3D indoor maps. Our evaluation among 12 indoor spaces and 10 users showcased the overall performance and usability of FlyMeThrough, and also revealed the potential future impact and application scenarios.

\subsection{Application Scenarios}
FlyMeThrough demonstrates a wide range of promising application scenarios that can assist building managers, occupants, and even first-time visitors in knowing, evaluating, and inspecting indoor spaces. The tool supports tasks such as navigation, locating rooms, and identifying smaller indoor features—from doors and stairs to electrical panels and AEDs. This versatility highlights its strong potential for broader adoption, especially given FlyMeThrough’s support for customizable mapping. With a quick annotation session lasting just a few minutes, users can generate tailored indoor maps to fit specific needs, including multiple versions for different use cases. The system also supports annotation of smaller, flexible, and temporary structures—such as furniture, cargo boxes, and exhibition setups—which are often omitted from traditional architectural maps. This enables the creation of detailed management maps for events, exhibitions, and furniture logistics. In addition, the high-resolution 3D maps provide rich spatial cues that can benefit individuals with accessibility needs, allowing them to better understand environmental conditions and plan their movements based on specific accessibility considerations.


\subsection{End-user Interface}
At this stage, we have implemented a web interface (\autoref{fig:review}) to showcase the final 3D map. In future work, we plan to develop additional UIs \xia{that reflect user study feedback and are} tailored to different user needs. For example, a navigation interface could better support spatial movement and route planning to specific indoor facilities; a facility management interface could focus on accessing and managing building safety information and equipment; and a mobile interface could support real-time user localization by matching camera feeds with the 3D indoor map to enable live navigation assistance. We also envision accessibility-focused interfaces that evaluate and visualize spatial accessibility based on individual user requirements. \xia{Furthermore, we expect to further automate the annotation process by incorporating confirmations of open-vocabulary object detection results.} Informed by our user study findings and the rich capabilities of our 3D mapping system, there is significant potential to explore a wide design space for future 3D mapping interfaces.

\subsection{Improve model quality \& reconstruction efficiency}
As discussed in the user study and limitations, the current 3D models generated using the commercial photogrammetry tool MetaShape \cite{over2021processing} can be improved in terms of quality. In future work, we aim to explore alternative 3D reconstruction methods that are fully open-source, better automated, and capable of producing higher-quality 3D maps.


\richard{Open-source methods are improving rapidly---for example, recent work such as \textit{VGGT-SLAM} \cite{maggio2025vggt} already supports processing hundreds of frames efficiently. While our evaluations (Section~3.2) found that existing open-source solutions did not yet meet our quality and robustness needs, our pipeline’s modular design allows us to seamlessly integrate improved reconstruction algorithms as they become viable.}
Another promising direction is generative view synthesis \cite{weber2025fillerbuster}, which enables the synthesis of novel views from captured images. This capability can help enhance reconstruction quality in areas that are occluded or otherwise inaccessible to the drone, and potentially reduce the extent of capture required.

We plan to continue exploring these open-source alternatives and will consider transitioning to them as their performance becomes suitable for our use cases.





\subsection{Privacy Guidelines}
\xia{Our user study reveals both concerns and mitigation methods around privacy and confidentiality of the scanned indoor spaces. Building upon our findings and recognizing the importance of privacy in indoor space scanning processes \cite{285417,yang2025privacy,fathalizadeh2024indoor}, we propose the following privacy guidelines:}

\xia{\textbf{Spatial Exclusion}. Involve the key facility stakeholders (\textit{e.g.} building management, building tenants) to maintain an exclusion list where the drone fly should not cover. For example, private units, confidential offices and labs, restrooms, \textit{etc.} }

\xia{\textbf{Temporal Minimization}. Schedule flights outside peak occupancy windows and publish a time-bounded flight schedule at least 48 hours in advance so tenants can vacate or opt out. Inadvertent capture and acoustic disturbance can also be minimized.}

\xia{\textbf{Data Management}. The recorded data should be automatically filtered for personal identification information (\textit{e.g.} faces) and certain building information (\textit{e.g.} confidential text that need to be omitted) prior to further processing. Store unredacted footage on an encrypted, access-controlled server. Limit the access to the raw data and processed 3D mapping data based on the building type and mapping goals.}

\subsection{Automating Drone Flights}
In future work, we will explore the automation of indoor drone flights to create a fully automated data collection pipeline, enabling regular, automated scans. Existing research \cite{gao2023uav, li2018universal, zhang2024icon, maboudi2023review, gao2024hierarchical} has already explored autonomous path-planning for drones navigating indoor environments, and we plan to build upon them to explore the feasibility of automating commodity drones at the software level. Based on our user study findings, we anticipate no major bystander concerns with automated drone data collection. The primary challenges lie in ensuring appropriate scanning coverage and timing, as well as effective obstacle avoidance to guarantee flight safety. To address these concerns, we propose methods such as setting no-fly zones within indoor spaces and detecting when the environment is too congested for safe flight, thereby mitigating potential risks.



\subsection{Comparing with commercial and LiDAR-based systems}
\richard{While direct empirical comparisons with commercial systems such as Matterport or LiDAR-based drones were not feasible due to proprietary restrictions and hardware availability, we examined publicly available documentation and prior studies to approximate how our method compares to existing alternatives.}

\richard{Efficiency emerged as a key advantage of our approach. Matterport and terrestrial laser scanning (TLS) systems typically require 30–90 minutes to scan a 100 m² indoor space \cite{chen2018accuracy}, whereas our system captured spaces more than 20 times larger in less than 10 minutes (~\autoref{tab:space_summary}), demonstrating an improvement in acquisition speed. This efficiency, combined with the ability to handle large-scale and complex indoor environments without additional hardware, highlights the practical benefits of our pipeline in time-sensitive or resource-constrained scenarios. Furthermore, among all methods tested in our experiments, our pipeline was the only one capable of producing reliable reconstructions across such large spaces, indicating not only efficiency but also robustness.}

\richard{We acknowledge that this efficiency comes at a cost to reconstruction quality. When comparing the reconstruction results with the raw input images, our average peak signal-to-noise ratio (PSNR)\cite{hore2010image} 
was approximately 10 dB, which is lower than the 15–23 dB reported by Burde \textit{et al.} \cite{burde2025comparative}
for 11 other methods on smaller-scale indoor scenes. This reflects a trade-off between speed and geometric fidelity that we aim to address in future work.}

\richard{It is important to note that these comparisons are approximate, as the methods differ in goals, assumptions, and testing environments. Nonetheless, they offer a high-level perspective on the strengths and limitations of our system in comparison to the broader landscape of indoor mapping technologies.}

\section{Conclusion}
In this paper, we present \textit{FlyMeThrough}, a drone-based indoor mapping system that leverages RGB-only drone footage and human-AI collaborative annotation to generate POI-infused 3D indoor maps. We evaluated the system in 12 indoor spaces of varying sizes, functions, and spatial layouts. A user study involving building managers and occupants demonstrated both high technical performance and strong usability of our system. Additionally, participants proposed a wide range of potential use cases, particularly in support of building management tasks.

\begin{acks}
This research was supported by the University of Washington's Center for Research and Education on Accessible Technology and Experiences (CREATE) and NSF grant \#2125087. We give special thanks to our study participants for their participation and feedback.
\end{acks}

\bibliographystyle{ACM-Reference-Format}
\bibliography{main}
\end{document}